\documentclass[sigplan,screen,pbalance]{acmart}

\providecommand{\correspondingauthor}{}

\usepackage{listings}
\usepackage{colortbl}
\usepackage{newfloat}
\usepackage{tikz}
\usetikzlibrary{fit, positioning, arrows.meta}
\usepackage{cleveref}
\usepackage[utf8]{inputenc}

\DeclareFloatingEnvironment[fileext=lol, listname={List of Listings},
  name=Listing, placement=tbp]{listing}
\crefname{listing}{Listing}{Listings}
\Crefname{listing}{Listing}{Listings}

\newcommand{\zjit}{ZJIT}

\newcommand{\rewrite}[2]{\mbox{$#1 \rightarrow #2$}}

\definecolor{synkeyword}{HTML}{D73A49}
\definecolor{synstring}{HTML}{298E0D}
\definecolor{syncomment}{HTML}{8A8A8A}
\definecolor{synfunction}{HTML}{4B69C6}

\lstdefinestyle{plain}{language={}}

\newenvironment{codeblock}
  {\par\addvspace{0.4\baselineskip}}
  {\par\addvspace{0.4\baselineskip}}

\newcommand{\codegap}{\vspace{0.8\baselineskip}}

\definecolor{cellgrey}{RGB}{200,200,200}   
\definecolor{cellghost}{RGB}{227,227,227}  
\definecolor{faintgrey}{RGB}{140,140,140}  
\definecolor{ghosttext}{RGB}{158,158,158}  

\newcommand{\epescapepanel}[1]{%
\begin{tikzpicture}[line width=0.5pt,
    cell/.style={draw, minimum size=0.6cm, inner sep=1pt, outer sep=0pt,
                 font=\small\ttfamily},
    proccell/.style={draw, minimum width=0.85cm, minimum height=0.42cm,
                     inner sep=1pt, outer sep=0pt, font=\small\ttfamily},
    idx/.style={font=\scriptsize, inner sep=0pt, outer sep=0pt},
    arr/.style={-{Stealth[length=3.5pt, width=3pt]}}]
  \ifnum#1=1
    \tikzset{scell/.style={cell, draw=faintgrey, text=ghosttext},
             sfill/.style={fill=cellghost},
             senvbox/.style={draw=faintgrey, densely dotted},
             stext/.style={text=faintgrey}}
  \else
    \tikzset{scell/.style={cell}, sfill/.style={fill=cellgrey},
             senvbox/.style={draw}, stext/.style={}}
  \fi
  \node[scell]        (s0) at (0.3, 0.3) {};
  \node[scell, sfill] (s1) at (0.9, 0.3) {a};
  \node[scell, sfill] (s2) at (1.5, 0.3) {b};
  \node[idx, stext, anchor=south] (sl0) at (0.3, 0.72) {0};
  \node[idx, stext, anchor=south]       at (0.9, 0.72) {1};
  \node[idx, stext, anchor=south] (sl2) at (1.5, 0.72) {2};
  \node[senvbox, inner sep=0.12cm, fit=(s0)(s2)(sl0)(sl2)] (senv) {};
  \node[idx, stext, anchor=south] (senvlbl) at ([yshift=0.06cm]senv.north)
    {environment};
  \node[draw, densely dotted, inner sep=0.12cm, fit=(senv)(senvlbl)] (vm) {};
  \node[idx, anchor=south] at ([yshift=0.06cm]vm.north) {VM Stack};
  \node[draw, minimum width=1.4cm, minimum height=1.1cm, inner sep=0pt,
        left=0.5cm of vm] (cfp) {};
  \node[idx, anchor=north] at ([yshift=-0.06cm]cfp.north) {CFP};
  \node[draw, minimum width=1.16cm, minimum height=0.46cm, inner sep=0pt,
        anchor=south, font=\scriptsize] (ep) at ([yshift=0.12cm]cfp.south) {EP};
  \coordinate (bus) at ([yshift=-0.35cm]vm.south);
  \ifnum#1=1
    \def\panellabel{(b) after escape}
    \coordinate (hx) at (vm.east |- 0,0);
    \coordinate (hbase) at ([xshift=1.5cm]hx);
    \node[cell]                (h0) at ([shift={(0.3,0.3)}]hbase) {};
    \node[cell, fill=cellgrey] (h1) at ([shift={(0.9,0.3)}]hbase) {a};
    \node[cell, fill=cellgrey] (h2) at ([shift={(1.5,0.3)}]hbase) {b};
    \node[idx, anchor=south] (hl0) at ([shift={(0.3,0.72)}]hbase) {0};
    \node[idx, anchor=south]       at ([shift={(0.9,0.72)}]hbase) {1};
    \node[idx, anchor=south] (hl2) at ([shift={(1.5,0.72)}]hbase) {2};
    \node[draw, inner sep=0.12cm, fit=(h0)(h2)(hl0)(hl2)] (henv) {};
    \node[idx, anchor=south] (henvlbl) at ([yshift=0.06cm]henv.north)
      {environment};
    \node[draw, densely dotted, inner sep=0.12cm, fit=(henv)(henvlbl)] (heap) {};
    \node[idx, anchor=south] at ([yshift=0.06cm]heap.north) {heap};
    \draw[arr] (senv.east) -- node[idx, above=0.1cm, text=red,
      font=\scriptsize\bfseries] {copied!} (henv.west);
    \node[proccell, anchor=south west] (p0)
      at ([shift={(0.15,-1.8)}]h0.south west) {code};
    \node[proccell, anchor=west] (p1) at (p0.east) {env};
    \draw ([shift={(-0.12,-0.12)}]p0.south west)
      rectangle ([shift={(0.12,0.42)}]p1.north east);
    \node[idx, font=\scriptsize\ttfamily, anchor=north west]
      at ([shift={(-0.04,0.37)}]p0.north west) {Proc};
    \draw[arr] (p1.north) -- (p1.north |- henv.south);
    \draw[arr] (ep.south) -- (ep.south |- bus) -- (h0.south |- bus) -- (h0.south);
  \else
    \def\panellabel{(a) before escape}
    \draw[arr] (ep.south) -- (ep.south |- bus) -- (s0.south |- bus) -- (s0.south);
  \fi
  \node[idx, anchor=north] at ([yshift=-0.15cm]current bounding box.south)
    {\panellabel};
\end{tikzpicture}}

\setcopyright{cc}
\setcctype{by-nc-nd}
\acmDOI{10.1145/3840562.3844969}
\acmYear{2026}
\copyrightyear{2026}
\acmISBN{979-8-4007-2932-4/2026/10}
\acmConference[VMIL '26]{Proceedings of the 18th ACM SIGPLAN International Workshop on Virtual Machines and Intermediate Languages}{October 4--9, 2026}{Oakland, CA, USA}
\acmBooktitle{Proceedings of the 18th ACM SIGPLAN International Workshop on Virtual Machines and Intermediate Languages (VMIL '26), October 4--9, 2026, Oakland, CA, USA}
\acmSubmissionID{splashws26vmilmain-p43-p}
\received{2026-08-15}
\received[accepted]{2026-08-27}

\begin{document}

\title{Support Local Variables}

\author{Maxwell Bernstein}
\correspondingauthor
\orcid{0000-0003-3130-7059}
\affiliation{%
  \institution{Shopify}
  \city{Boston}
  \country{USA}
}
\email{acm@bernsteinbear.com}

\author{Takashi Kokubun}
\orcid{0009-0007-2567-3016}
\affiliation{%
  \institution{Shopify}
  \city{Cupertino}
  \country{USA}
}
\email{takashi.kokubun@shopify.com}

\author{Aaron Patterson}
\orcid{0009-0003-6079-9125}
\affiliation{%
  \institution{Shopify}
  \city{Seattle}
  \country{USA}
}
\email{aaron.patterson@gmail.com}

\author{Si Xing (Alan) Wu}
\orcid{0009-0003-8399-2251}
\affiliation{%
  \institution{Shopify}
  \city{Ottawa}
  \country{Canada}
}
\email{paper@alanwu.email}

\author{Kevin Menard}
\orcid{0009-0005-4551-5555}
\affiliation{%
  \institution{Shopify}
  \city{Boston}
  \country{USA}
}
\email{acm@kevin.nirvdrum.com}

\begin{abstract}
Ruby is a dynamically typed and object-oriented programming language. Its
primary implementation, CRuby, contains a bytecode virtual machine and a
mature lazy basic block versioning (LBBV) just-in-time (JIT) compiler
called YJIT.

In order to both implement more advanced optimizations than YJIT supports
and also encourage more outside contributions, we present a new
method-based JIT called
ZJIT. Like YJIT, ZJIT compiles from bytecode to machine code. Unlike
YJIT, ZJIT has multiple global and local
optimization passes.

ZJIT's high-level intermediate representation is in static single
assignment (SSA) form. In order to optimize Ruby's local variables, ZJIT
lifts local variables into SSA values. This is a departure from how
other Ruby compilers handle locals: other JIT compilers either leave local
variables as memory loads and stores or do advanced partial evaluation to
recover SSA values from memory.

While implementing locals, we (re-)discovered what features make local
variables in Ruby especially challenging to compile correctly and
efficiently. We demonstrate these features and illustrate how we solved
these problems in ZJIT.
\end{abstract}



\maketitle

\section{Introduction}

Just-in-time compilers accelerate programs either immediately before or
interleaved with execution. They are commonly used to efficiently implement
dynamic languages because very little information is available ahead of time.

JIT compilers are often a second implementation of the source language. In
order to execute code faster, they implement the language differently
than the existing implementation. They may even represent objects differently,
including the method call frames.

One oft-used language feature is local variables, which are used for storing
temporary values in a method call frame. Locals are
so common that most developers will never take pause before using them.
However, Ruby semantics for local variables have presented many challenges for
our new JIT compiler, \zjit{}.

Unlike the existing compiler, YJIT, \zjit{} does not flush and re-load local
variables from the VM frame. Instead, \zjit{} lifts local variables to SSA values
and uses speculation and deoptimization to recover from mispredictions. For the
last year, we have incrementally discovered more semantic dark corners that we
need to faithfully implement in our new compiler for compatibility.

We implement this compiler in the context of CRuby's bytecode interpreter, YARV.
Both are described in \Cref{sec:background}. It is compiled and integrated much the same
way as YJIT. It has similar compatibility as YJIT; we pass the Ruby test suite
and can successfully run many large web applications, with only a handful of
known bugs.

This speculative approach for local variables is novel in Ruby compilers: Ruby
allows so much reflection that its other major JIT-compiled implementations
(YJIT, JRuby, and TruffleRuby) use different approaches, described in \Cref{sec:related}.
Our SSA lifting approach has two major motivations.

First, it helps us better optimize our high-level intermediate representation:
for example, delete redundant type checks, eliminate memory loads and stores,
and more effectively apply value numbering.

Second, because we are grafting a new compiler onto an existing runtime, we
must work within the context of a large mass of existing C library code, little
of which was designed with JIT compilation in mind. Therefore, a
partial-evaluation-based approach would recover minimal information about local
variables. We have found it reasonable to add callback ``hooks'' to
deoptimization points in the existing C codebase.

This paper will discuss existing implementation background in \Cref{sec:background},
introduce \zjit{} and its architecture in \Cref{sec:architecture}, and present challenges
for implementing local variables with Ruby-specific semantics in \zjit{} in
\Cref{sec:locals}. Then it will present its current (and projected future) solutions to
these problems in \Cref{sec:optimizing-locals}.

\section{Background}
\label{sec:background}

\subsection{The Ruby Programming Language}

Ruby has features that make the optimization of the language challenging.
Every value in a Ruby program is an object; there are no primitive types
for integers, for example. Nearly every operation on an object is a method call, and
every method can be redefined at run time. Ruby has \texttt{eval} and other APIs that
allow programmatic reflection, including access to local variables, which will be discussed in \Cref{sec:locals}.

\subsection{The CRuby Virtual Machine}

CRuby is the reference implementation of Ruby. CRuby is implemented
as a stack-based interpreter of the YARV bytecode instructions~\cite{sasada2005}.
CRuby implements a significant portion of its built-in libraries in C functions,
which make it difficult for compilers to understand the behavior of a program
from the bytecode. More details are described in recent YJIT papers~\cite{chevalierboisvert2021, chevalierboisvert2023}.

\subsection{YJIT}

YJIT~\cite{chevalierboisvert2021, chevalierboisvert2023} is a lazy basic block versioning
(LBBV)~\cite{chevalierboisvert2015, chevalierboisvert2016} just-in-time (JIT) compiler,
which was retrofitted onto the CRuby virtual machine. YJIT compiles the YARV bytecode
into arm64 or x86\_64 machine code. LBBV compiles one basic block at a time and
propagates the information of types and frames to succeeding basic blocks. YJIT encodes
such block-specific contexts in a compact representation called Context~\cite{chevalierboisvert2023}.
To minimize the memory required for Context, YJIT does not encode values observed in blocks.
This design makes it challenging for YJIT to implement cross-block optimizations that need
to know more than types and frames, such as constant folding.

\section{Architecture of \zjit{}}
\label{sec:architecture}

\zjit{} is a new method-based JIT compiler for CRuby. Unlike YJIT, which compiles
one basic block at a time, \zjit{} compiles a whole method and any inlined callees
at once, enabling more advanced optimizations than YJIT supports. We also aim
to encourage more outside contributions by choosing a traditional SSA-based
compiler design.

\subsection{Profiling}
\label{sec:profiling}

Ruby bytecode carries no type information, and nearly every optimization
\zjit{} performs depends on knowing the classes and shapes of the
values that flow through a method.

YJIT learns type information with lazy
basic block versioning~\cite{chevalierboisvert2015, chevalierboisvert2016}
by interleaving compilation and execution. Right before a basic block
executes, the compiler can inspect the actual operand values sitting on the VM stack
at compile-time. \zjit{} does not have this luxury; it
compiles an entire method at once, including paths that have not yet executed,
so it must gather type information from the interpreter.

Profiling is driven by the per-method call counter that decides
when to compile. A method starts getting profiled after P calls (25 by default) and starts getting compiled after C further calls (5 by default).
\zjit{} profiles by rewriting
the method's bytecode in place, replacing each instruction it wants to
profile with a profiling variant of that instruction. A profiling variant
records the operand types of each instruction in a side-table and then executes the original
instruction. When profiling is done, profiling instructions get rewritten to
their original opcodes.

The side table is a feedback vector~\cite{hoelzle1994}
that records the distribution of up to four observed types. It resembles
the type feedback structure in S6, a JIT compiler for CPython~\cite{s6}.

This design differs from quickening~\cite{brunthaler2010}, where observed
runtime behavior is encoded by rewriting bytecode into
specialized instructions.

\subsection{Optimization}
\label{sec:optimization}

After profiling bytecode, we construct an SSA-based high-level intermediate
representation, HIR. We construct this by performing an abstract
interpretation over CRuby bytecode in several phases:

\begin{enumerate}
\item Discover bytecode basic block boundaries in one linear pass by marking which opcodes
  are jump targets.
\item Create an entry HIR basic block for the interpreter and one additional JIT
  entry basic block for each potential arity of optional parameters passed.
\item Push the entry bytecode basic block onto a worklist.
\item Using the worklist and a visited set, walk the bytecode's basic blocks until
  the worklist is empty, transforming each bytecode basic block into HIR.
\end{enumerate}

While building HIR from one bytecode basic block, we maintain a compile-time virtual stack and
compile-time virtual locals array that mirror the interpreter's run-time data
structures. When an opcode handler would, in the interpreter, push to the stack, we
push to the virtual stack. When an opcode would dispatch to a runtime helper
(e.g. \texttt{rb\_ec\_ary\_new\_\allowbreak from\_\allowbreak values}), we create an HIR instruction (e.g.
\texttt{NewArray}). We show the correspondence between the interpreter's stack and
compiler's virtual stack in \Cref{lst:interp-ssa} interpreting/compiling the expression \texttt{1 + 2}.

Our HIR is represented as instructions that refer to their operands with
pointers as in the HotSpot client compiler~\cite{kotzmann2008}, so entries in these
compile-time data structures are instruction pointers.

\newsavebox{\interpbox}
\newsavebox{\ssabox}
\begin{lrbox}{\interpbox}
\begin{minipage}[t]{0.40\columnwidth}
\begin{lstlisting}[style=plain, xleftmargin=0pt, basicstyle=\ttfamily\scriptsize]
; [] :: Stack[VALUE]
pushconst:
  PUSH(1)
; [1]
pushconst:
  PUSH(2)
; [1, 2]
send:
  r = POP()
  l = POP()
  x = send(l, :+, r)
; [3]
\end{lstlisting}
\end{minipage}
\end{lrbox}
\begin{lrbox}{\ssabox}
\begin{minipage}[t]{0.50\columnwidth}
\begin{lstlisting}[style=plain, xleftmargin=0pt, basicstyle=\ttfamily\scriptsize]
; [] :: Stack[Insn*]
pushconst:
  v0 = emit(Const 1)
; [v0]
pushconst:
  v1 = emit(Const 2)
; [v0, v1]
send:
  v2 = emit(
    Send v0, :+, v1
  )
; [v2]
\end{lstlisting}
\end{minipage}
\end{lrbox}

\begin{listing}[t]
\centering
\scriptsize
\setlength{\arrayrulewidth}{0.5pt}
\arrayrulecolor{gray}
\begin{tabular}{@{}l@{\hspace{0.8em}}|@{\hspace{0.8em}}l@{}}
\textbf{Bytecode interpreter} & \textbf{SSA} \\[0.3em]
\usebox{\interpbox} & \usebox{\ssabox} \\
\end{tabular}
\arrayrulecolor{black}
\caption{Bytecode interpreter operations vs SSA construction from bytecode.}
\label{lst:interp-ssa}
\end{listing}

To construct HIR for the entire control-flow graph (CFG), we construct \emph{maximal
SSA} following Appel's ``really crude approach'' of generating a
$\phi$\footnote{Really, a basic block parameter.} for every stack value and
local variable at basic block boundaries~\cite{appel_ssa_1998}.

While building HIR, we specialize instance variable reads and writes using
profiles and the interpreter's \emph{shape} machinery~\cite{chambers1989}. After building
HIR, we enter the optimizer.

Our optimizer is inliner-driven~\cite{prokopec2019}, using inlining heuristics to
explore a subset of the program's call tree. We iteratively inline and
optimize the resulting code until we reach a configurable stopping condition:

\begin{enumerate}
\item Ingest interpreter profiles (see \Cref{sec:profiling}). Specialize method lookups and
  the well-known builtins, taking care to insert PatchPoint instructions
  whose code location is rewritten to a side-exit code when the methods change
  (see \Cref{sec:deoptimization}).
\item Perform an inlining step. Use a worklist to do a breadth-first search of the
  available inlinable call-sites.
\item Minimize SSA~\cite{aycockSimpleGenerationStatic2000, braun2013}.
\item Perform store-to-load forwarding and redundant store elimination.
\item Fold constants (\rewrite{x \times 0}{0}), strength reduce operations
  (\rewrite{x / 8}{x \gg 3}), and eliminate redundant guards.
\item Collapse linked lists of branches in the CFG.
\item Remove redundant PatchPoint instructions and interrupt checks.
\item Eliminate dead code.
\end{enumerate}

In between each pass, we perform a flow typing over the CFG, iterating SSA
values' types to fixpoint.

\subsection{Code Generation}

We lower high-level IR to low-level IR by mapping each HIR basic block to one or more
LIR basic blocks, and each HIR SSA value to one LIR operand. We implement many HIR
operations in-line in LIR, and implement others by dispatching to runtime
helper functions written in C. After construction, we allocate registers
in-place.

In preparation for register allocation, we schedule and number LIR
instructions. Then, for each virtual register, we compute intervals with
lifetime holes and perform linear scan register allocation using a partial
implementation of Wimmer's 2005 and 2010 algorithms~\cite{wimmer2005, wimmer2010}.

After assigning register and spill slots to intervals, we lower the LIR to
arm64 or x86\_64 machine code.

Each block of machine code has multiple entries to the same function body:
an interpreter entry and JIT call entries. When the interpreter
dispatches JIT code, it calls into a globally-shared trampoline
that sets up a set of callee-saved registers that are shared among
all JIT code, and then jumps to the method-specific interpreter entry,
which loads method parameters from the VM stack into registers or
native stack slots used by the function body. JIT call entries,
on the other hand, load method parameters based on the C calling
convention, which allows JIT code to call them as if they were
a regular C function.

\subsection{Deoptimization}
\label{sec:deoptimization}

In Ruby, a single variable
can have an object of any class, which can make the destination of a method
dispatch on it unpredictable. When the type of a method receiver cannot be
inferred from HIR, \zjit{} speculates that it will be one of the types
observed in profiles and inserts guards that check the type at run time.
When the guard fails, it side-exits from the compiled code~\cite{fink2003, steiner2007}, delegating the
execution of that instruction and the rest of that method's activation to the
interpreter.

Ruby also has meta-programming features. For instance, methods and constants
can be redefined at any time. \zjit{} speculates that methods and constants
referenced by JIT code will not be redefined. Instead of inserting a run-time
check for it, \zjit{} inserts PatchPoint instructions, which are no-op in code and
only remember the address of locations where such speculations are made, and
rewrites them to a side-exit code when such redefinition indeed happens.

In CRuby, many built-in methods and third-party libraries are written in C,
which \zjit{} cannot inline and see through in HIR, and they may raise an
exception that is implemented using longjmp. When it does a longjmp to
CRuby's exception handler, it expires the caller native frames for JIT code.
\zjit{} does deoptimization~\cite{deutsch1984, hoelzle1992, duboscq2013} before such
a longjmp to copy the stack operands that will be used by the bytecode from
the native stack before they expire. On every safepoint where \zjit{} needs to
prepare for deoptimization, it allocates deoptimization metadata on the heap
at compile-time, which encodes the relationship between physical storage
locations
and the VM's frame state as a stack map, and writes a single pointer to it
at run-time.

\subsection{Recompilation}

When a method or a constant is redefined and its associated PatchPoints are
rewritten to a side exit, \zjit{} invalidates entries to the JIT code that had
the PatchPoints and resets the call counter of the invalidated methods.
The interpreter will re-profile the method for a configured number of calls,
and when the call counter reaches the threshold again, the method will be
recompiled. A method is allowed to have up to four versions by default.

On the other hand, when JIT code side-exits to the interpreter due to a type
guard failure, it does not immediately invalidate entries to the JIT code.
The execution of future method calls will still start in the JIT code. However,
after the JIT code side-exits, the interpreter will profile the instruction that
exited as well as the rest of the instructions in the method. After a configured
number of profiles are collected on the exited instruction, \zjit{} recompiles
a new version of the method, redirecting any entries from the previous version
to the new version.

\section{Local Variables}
\label{sec:locals}

Now that we have described \zjit{}, we will describe the semantics of local
variables in the Ruby language.

\subsection{The Basics}

Local variables may be assigned and read in a method. They may be assigned many
times. Each method defined using the \texttt{def} keyword creates a root local
variable environment; locals must be defined in the method in order to be
usable in the method.

\begin{codeblock}
\begin{lstlisting}[language=Ruby]
def foo
  a = 1; a = 2; a  # 2
end

a = 10
def bar
  a  # undefined local variable or method 'a'
end
\end{lstlisting}
\end{codeblock}

Locals are method-scoped and have an initial value of \texttt{nil}. In the interpreter, setting up a frame
involves filling \texttt{nil}s in the VM frame.

\begin{codeblock}
\begin{lstlisting}[language=Ruby]
def maybe(cond)
  if cond
    x = 3
  end
  x  # 3 or nil
end
\end{lstlisting}
\end{codeblock}

Locals may be used before they are defined. However, because method calls and
local variables have syntactic overlap, referring to an identifier may either
cause a method call or load a local variable.

A reference to an identifier is resolved at parse-time to be either a
local variable reference or a method call with no arguments. Assigning to a
local variable causes references syntactically ``further down'' to resolve as local
variable reads.

\begin{codeblock}
\begin{lstlisting}[language=Ruby]
def me = puts("me")
def call_me_maybe
  me; me = 1; me  # "me" followed by returning 1
end
\end{lstlisting}
\end{codeblock}

Local variables are typically stored on the interpreter stack, but can ``escape''
to the heap when captured by a first-class object. We describe this in \Cref{sec:proc}.

\subsection{Blocks}

Ruby has a construct called a \emph{block}. A block is similar to a \texttt{lambda} in
Scheme in that it can refer to and mutate its enclosing environment. For
example, in the following code, \texttt{count} lives at level 0 and the block refers
to it from level 1, mutating the enclosing environment:

\begin{codeblock}
\begin{lstlisting}[language=Ruby]
count = 0
[1, 2, 3].each { count += 1 }
# count is 3
\end{lstlisting}
\end{codeblock}

These block environments can be nested to an arbitrary depth. For example, in
the following code, the innermost block is able to reach up two levels to
modify \texttt{count}.

\begin{codeblock}
\begin{lstlisting}[language=Ruby]
count = 0
[1, 2, 3].each {
  [4, 5, 6].each { count += 1 }
}
# count is 9
\end{lstlisting}
\end{codeblock}

A method can take a block as a parameter, as with the \texttt{each} method above:

\begin{codeblock}
\begin{lstlisting}[language=Ruby]
def calls_block(&block)
  yield  # or `block.call`
end
\end{lstlisting}
\end{codeblock}

The block parameter is a simple reference to the block's code; it can be invoked without capturing the environment.
Other uses trigger a transformation into a \texttt{Proc}, which pairs the code with a captured environment.

\subsection{Procs}
\label{sec:proc}

When referred to by name\footnote{Most of the time. Some special cases do
not, for the purposes of optimization.}, a block gets transformed into a \texttt{Proc} object:

\begin{codeblock}
\begin{lstlisting}[language=Ruby]
def calls_block(&block) = block
calls_block { }  # => #<Proc: ...>
\end{lstlisting}
\end{codeblock}

Proc objects are a tuple of $(\textsf{code}, \textsf{env})$. When the interpreter
creates a Proc object, it moves the referenced environment to the heap, which
we call ``environment escape''. The
Proc's captured environment can then be arbitrarily read and written by any code with a reference to it. This means that the local variable \texttt{a} can be modified even though
it is not visibly written (or referenced at all!) in the block:

\begin{codeblock}
\begin{lstlisting}[language=Ruby]
def calls_block(&block) = block
a = 1
p = calls_block { }
p.binding.local_variable_set(:a, 2)
a  # 2
\end{lstlisting}
\end{codeblock}

This presents an optimization challenge for JIT compiler authors. If escaping
the environment and arbitrarily modifying it were a common operation, our
optimistic approach to local variables would not be as effective and we would
need to run slow deoptimization code; we assume such escape operations are infrequent.
Fortunately, environment escaping and writing to local variables via a bound
environment does not happen frequently. In a benchmark of a
large real-world Rails application, we observe the environment escaping roughly 35 times
per web request. Compared to other ``slow path'' events, which can occur hundreds
of thousands of times, this is relatively rare.

\section{Optimizing Locals}
\label{sec:optimizing-locals}

We are therefore motivated to optimize the ``fast path'': local variables that
remain on the stack. We want to assign HIR SSA values to every local variable,
as we do with every stack entry (instead of, say, emitting frequent load and store instructions). This not only avoids reading and writing
memory but also allows us to compute, sparsely store, and share type
information about each local variable.

We choose this path instead of partial escape analysis~\cite{stadler2018} because our runtime is
predominantly written in C and the JIT compiler cannot reason about C code.
Partial escape analysis would therefore not effectively be able to bring locals into SSA.

So we lift locals to SSA values during SSA construction. Below is a sample
trace of constructing HIR from a Ruby method that initializes two locals \texttt{a}
and \texttt{b} and then constructs an array containing both. We use the FrameState
structure in our abstract interpretation of the bytecode to map local variables
to SSA values.

\begin{codeblock}
\begin{lstlisting}[language=Ruby]
def make_array
  a = 1; b = 2; [a, b]
end
\end{lstlisting}
\codegap
\begin{lstlisting}[style=plain]
FrameState { locals = { } }
v0 = Const nil
FrameState { locals = { a: v0 } }
v1 = Const nil
FrameState { locals = { a: v0, b: v1 } }
v2 = Const 1
FrameState { locals = { a: v2, b: v1 } }
v3 = Const 2
FrameState { locals = { a: v2, b: v3 } }
v4 = NewArray v2, v3
Return v4
\end{lstlisting}
\end{codeblock}

This FrameState approach is not only useful for doing into-SSA
translation; it also helps us build stack maps for exiting JIT code into
the interpreter (see \Cref{sec:deoptimization}).

\subsection{Writing to Enclosing Environments}
\label{sec:mem-slot-for-lexical-captures}

Writing to enclosing environments is rare. In an execution of a large Rails
benchmark, for example, we see 4 million reads from an enclosing environment
but only 300,000 writes to an enclosing environment. This motivates penalizing
writes in order to make reads fast.

\begin{codeblock}
\begin{lstlisting}[language=Ruby]
count = 0
[1, 2, 3].each { count += 1 }
\end{lstlisting}
\end{codeblock}

In CRuby, blocks are represented by their own instruction sequences instead of
coded in-line. In the above example, there are three instruction sequences of
note:

\begin{enumerate}
\item The ``main'' method containing \texttt{count = 0} and the call to \texttt{each}
\item The \texttt{each} method, which is originally defined in Ruby's built-in \texttt{Array}
  class
\item The block containing \texttt{count += 1}
\end{enumerate}

The ``main'' instruction sequence refers to the \texttt{each} method by name and refers
to the block by pointer. The ``main'' instruction sequence invokes \texttt{each},
passing a block containing the address to the block. The \texttt{each} instruction
sequence contains no pointer to the block and receives it as a sort of formal
parameter\footnote{In a special location on the VM frame called the ``block
handler''.}.

Due to this setup, all three instruction sequences are given their own
translation units unless the main instruction sequence inlines \texttt{each} and then
again inlines the block\footnote{Inlining blocks is not yet supported and is
left for future work.}.

Our current solution to supporting writes to local variables from within blocks
is to:

\begin{enumerate}
\item Flush the VM frame before a send to a method with a block.
\item Execute the send.
\item Re-load local variables that are syntactically written to within that block.
\end{enumerate}

This works most of the time, but hits three snags: inlining, \emph{environment
escape} and \emph{eval}. We will discuss the latter two in \Cref{sec:environment-escape} and
\Cref{sec:eval}.

Inlining presents a problem: if we inline the method and the method body does
not contain any code that would require a VM frame, we do not end up flushing
the VM frame. Then, since we re-load after the send, we re-load garbage data
from the VM frame.

Furthermore, our current solution is too eager to write to the VM frame: we
would prefer to lazily write to the frame only when needed.

Therefore, our next implementation will likely instead ensure frame consistency
another way. In order to make parent and child environments agree, we make
block-referenced locals live on the VM frame and get written to and read from
that location consistently.

This means that the Ruby local \texttt{i} doesn't get a single SSA value across
uses; each use gets its own instruction:

\begin{codeblock}
\begin{lstlisting}[style=plain]
  SetLocal level=0, i, 0
  [1, 2, 3].each {
      v0 = GetLocal level=1, i
      v1 = FixnumAdd v0, 1
      SetLocal level=1, i, v1
  }
\end{lstlisting}
\end{codeblock}

That solves that problem: \texttt{GetLocal} and \texttt{SetLocal} do the requisite amount of
pointer chasing to find the frame at different levels and read/write \texttt{i} to
memory.

\subsection{Handling Environment Escape}
\label{sec:environment-escape}

The interpreter models local variables using an environment pointer (EP) and a local
index. The environment pointer most commonly points to the VM stack. However,
the environment can ``escape'' and move to the heap. The interpreter therefore
loads the environment pointer from a VM frame (Control Frame Pointer, or ``CFP'') for every local operation.

\begin{codeblock}
\begin{lstlisting}[language=C, xleftmargin=0pt]
int just_a() {  // as it executes in the interpreter
  // a = 1
  ep = cfp->ep
  *(ep + a_offset) = 1
  // return a
  ep = cfp->ep
  v0 = *(ep + a_offset)
  return v0
}
\end{lstlisting}
\end{codeblock}

\begin{figure}
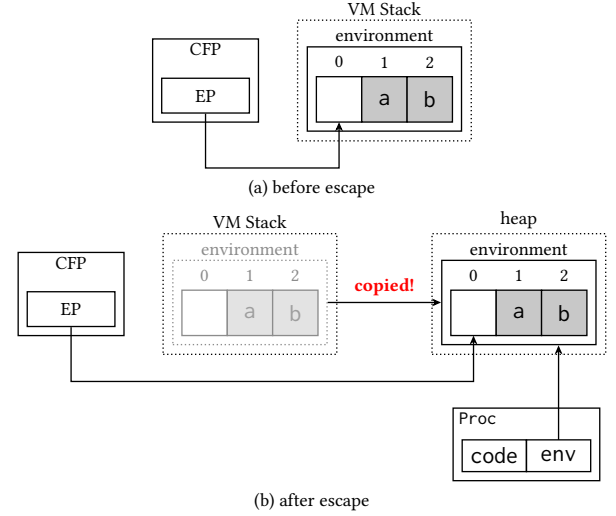

\centering
\epescapepanel{0}
\par\medskip
\epescapepanel{1}
\caption{Environment escape. In (a) the environment lives in the frame on the
  VM stack and \texttt{cfp->ep} points to it. When a \texttt{Proc} captures the
  environment (b), it is copied to the heap and \texttt{cfp->ep} is updated to
  point at the new location.}
\label{fig:ep-escape}
\end{figure}

\Cref{fig:ep-escape} illustrates the environment before and after it escapes.

As mentioned in \Cref{sec:proc}, blocks potentially capture all locals in the environment
into a first-class object. When the environment is in an object, the compiler
cannot predict the value of any local variable after a call to any code outside
the compilation unit; any code could have a reference to the environment object
and use it to modify the local. The creation of an environment object thus
invalidates the SSA form we produce for access to local variables not assigned
a memory slot through accesses lexically visible in the manner described in
\Cref{sec:mem-slot-for-lexical-captures}.

We perform SSA conversion assuming local variables are never modified through
environment objects and use code patching to exit to the interpreter in the
rare event that the environment escapes to the garbage collector managed heap.

In the following example, we put a constant for the read of \texttt{i} in the HIR
program, as the block passed to \texttt{blackbox} does not lexically modify the
variable. In case \texttt{blackbox} causes an environment escape (``EP escape''), we exit to the
interpreter before the potentially incorrect \texttt{Return} is reached.

\begin{codeblock}
\begin{lstlisting}[language=Ruby]
def access_across_call
  i = 2
  blackbox { }
  i
end
\end{lstlisting}
\codegap
\begin{lstlisting}[style=plain]
HIR:
  v0 = Const 2
  v1 = FrameState { locals = { i: v0 } }
  v2 = Send :blackbox, block={ }
  PatchPoint NoEPEscape { v1 }
  Return v0
\end{lstlisting}
\end{codeblock}

We describe details of the code patched into \texttt{PatchPoint NoEPEscape} in the
next section.

\subsection{Exiting through \texttt{NoEPEscape}}

Patching the \texttt{PatchPoint NoEPEscape} in the previous section transforms the
program to the equivalent of the following (though it is a quick machine code
patch, not a whole-method recompilation):

\begin{codeblock}
\begin{lstlisting}[style=plain]
HIR:
  v0 = Const 2
  v1 = FrameState { locals = { i: v0 } }
  v2 = Send :blackbox, block={ }
  SideExit NoEPEscape { v1 }
\end{lstlisting}
\end{codeblock}

The pre-patch code does not use a memory slot for local variable \texttt{i}, preferring to reify the
constant on-demand. However, the interpreter expects a memory slot for each local variable, so the
\texttt{SideExit} potentially needs to initialize memory.

All activations of the method share the newly patched JIT code. While the
activation in the call stack initiating the environment escape will reach the
\texttt{SideExit} with an escaped environment, other activations in different
execution threads may reach it with an environment that has not escaped.

If the \texttt{SideExit} sees an escaped environment, then the \texttt{Send} could have modified local variables
using an associated \texttt{Proc} object, invalidating the pre-send \texttt{FrameState} (\texttt{v1} in the example). The
exit should not write to memory in this case.

If the \texttt{SideExit} sees an environment that has not escaped, we want to infer from that the \texttt{Send}
has not written to the locals, that the pre-send \texttt{FrameState} is still accurate post-send, so the
exit should initialize memory. However, for this inference to be correct, there must not be any API
in the system which can modify local variables without also causing an environment escape. In the
next section, we describe handling of \texttt{eval}, which can modify local variables without causing an
environment escape.

\subsection{Handling eval}
\label{sec:eval}

The scheme described thus far does not consider dynamically generated code:

\begin{codeblock}
\begin{lstlisting}[language=Ruby]
i = 0
eval("i = 1")
puts i
\end{lstlisting}
\end{codeblock}

\texttt{eval} dynamically compiles new code that shares a VM frame with the
existing method, allowing the \texttt{eval}ed code to read and write local variables. When running in the
interpreter, \texttt{eval} does not necessarily cause the environment to escape; the interpreter has all
local variables in memory and \texttt{eval} knows the layout.

We hook the \texttt{eval} event in order to both reify the VM frame and so that after
the \texttt{eval}ed code finishes, the code will continue execution in the interpreter
instead of in JIT code. Having \texttt{eval} always perform environment escape when \zjit{}
is enabled allows reusing \texttt{PatchPoint NoEPEscape} for exiting from JIT code.

We currently do not attempt to make \texttt{eval} fast as it is generally absent from performance-sensitive
applications.

\subsection{Interactions with Stack Maps}

Currently, to allow the pre-existing routines for performing environment
escapes to find local variables, \zjit{} writes all locals to memory prior
to calling methods that potentially cause environment escape. Also, if the SSA values for
local variables survive the method call, the register allocator also preserves
the value in native stack memory. Our current implementation thus tends to write two
copies of each local variable to memory prior to making a method call.

As environment escapes are rare, we plan to use stack maps to describe
where the register allocator preserves SSA values for local variables, so a
maximum of one copy of each local variable is written to memory (if spilled by the register allocator). 
The environment escape routine will be modified to inspect the stack map
to find each local variable. It can then copy and permute the local variables
into the general in-memory environment storage format.

\section{Evaluation}

In this section, we present the preliminary evaluation of \zjit{} performance,
comparing it with other Ruby VMs.

\subsection{Methodology}

We have used 4 benchmarks from the ruby-bench\footnote{\url{https://github.com/ruby/ruby-bench}, formerly known as yjit-bench~\cite{chevalierboisvert2021, chevalierboisvert2023}}
benchmark suite, focusing on evaluating the performance of local variables, method calls, and Rails-based web applications.

The \texttt{30k\_variables} and \texttt{30k\_methods} benchmarks are written by us, which stress-test local variables and method calls.
\texttt{30k\_variables} calls 300 methods that use 100 local variables where most assignments are variable-to-variable copies.
\texttt{30k\_methods} has 30,000 methods that only make a single method call to each other.

The \texttt{erubi-rails} and \texttt{railsbench} benchmarks are Rails-based web application benchmarks, which were included in the evaluation of \cite{chevalierboisvert2023}.
\texttt{erubi-rails} uses Rails template rendering to render a view from Discourse, a real-world Rails application.
\texttt{railsbench} generates HTTP responses, querying entries from a SQLite database.

Experiments on \zjit{}, YJIT, and the CRuby interpreter were based on CRuby version 4.1.0dev, CRuby commit hash c1e6dbd8c0\footnote{\url{https://github.com/ruby/ruby}}.
Experiments with JRuby used version 10.1.1.0. Experiments with TruffleRuby used version 34.0.1.
As \cite{chevalierboisvert2023} did, JRuby enabled \texttt{invokedynamic} as well as the parallel GC, and TruffleRuby enabled the JVM mode.

We used a machine that runs Ubuntu Linux 24.04.4 on an Intel Core Ultra 7 270K Plus processor with 32GiB of RAM and CPU frequency scaling disabled.
Each benchmark was run for 1,000 iterations for each Ruby VM and the first half of recorded iterations was discarded as warm-up time.

\subsection{Performance}

\begin{figure}[t]
\centering
\includegraphics[width=\columnwidth]{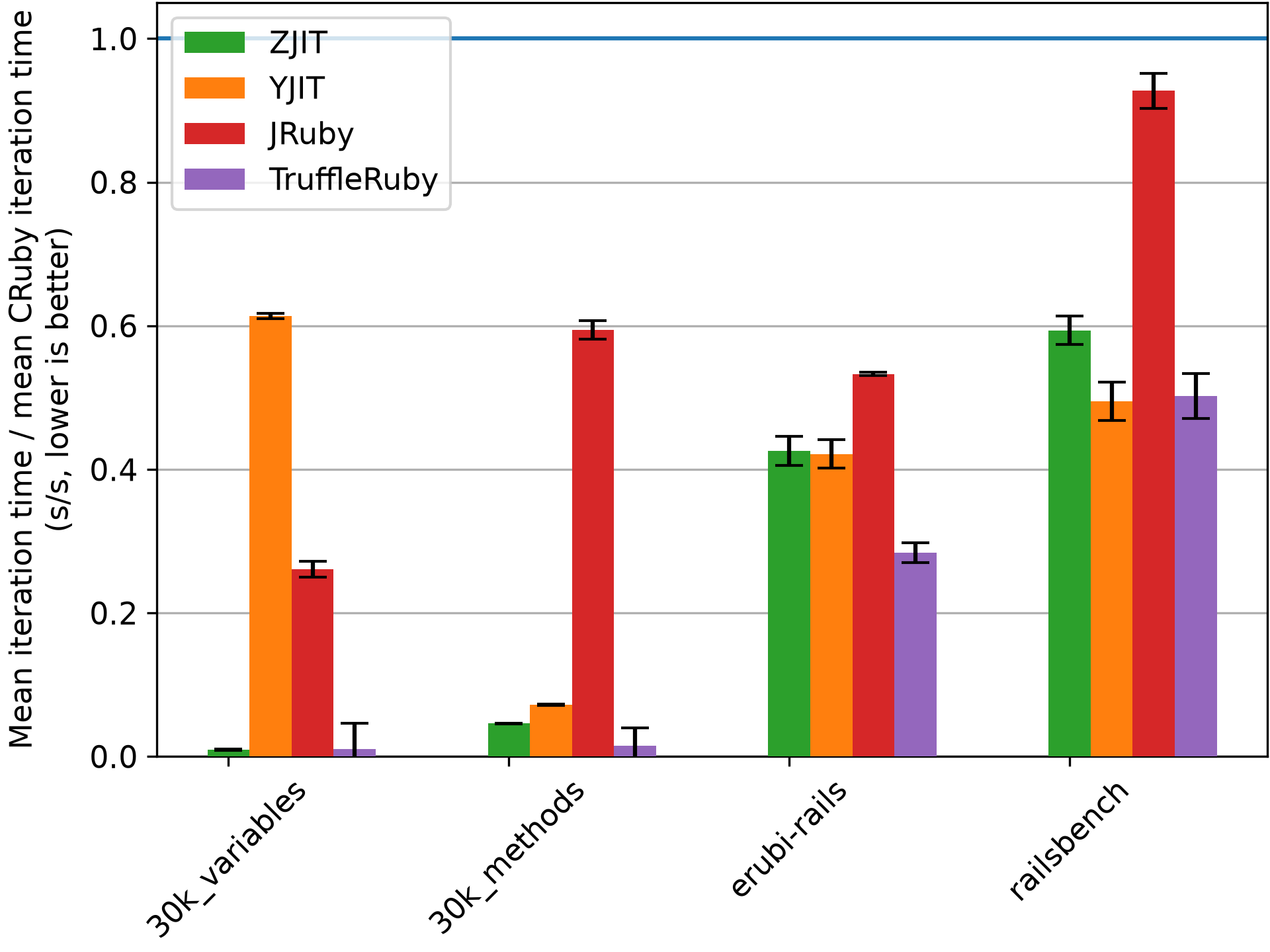}
\caption{Mean time per iteration of each Ruby implementation relative to the CRuby interpreter (lower is better). Error bars indicate one standard deviation.}
\label{fig:bars-speedup}
\end{figure}

\Cref{fig:bars-speedup} shows the mean time per iteration of \zjit{}, YJIT, JRuby, and TruffleRuby, relative to the CRuby interpreter.

Since \zjit{} lifts local variables into SSA values, variable-to-variable copies in \texttt{30k\_variables} incur no cost for \zjit{},
making it perform as well as TruffleRuby with low variance. YJIT allocates up to 5 registers for 5 fixed local variables,
but the remaining 95 local variables operate on memory, which makes YJIT much slower than \zjit{}.

On method calls, which are also safepoints, \zjit{} writes a single pointer to deoptimization metadata and skips writing
some of the VM frame fields, whether the call is inlined or not. YJIT doesn't have that optimization, so method calls
run faster on \zjit{} than on YJIT in \texttt{30k\_methods}. \zjit{} still writes some of the frame fields eagerly, which is left
for future work, and that makes \zjit{} slower than TruffleRuby.

\texttt{erubi-rails} and \texttt{railsbench} show that \zjit{}'s performance on real-world workloads is comparable to YJIT's while
still slightly lagging behind. The current performance difference between YJIT and \zjit{} comes from the fact that \zjit{}
lacks support for specializing some of the complex argument passing features in Ruby, such as array splat arguments,
forcing such calls to use the interpreter's C functions. They are also left for future work.

\section{Related Work}
\label{sec:related}

It is instructive to see how other language VMs handle the potential escaping of locals that cannot be statically determined to remain local. Ruby has several implementations in active development, each choosing a different way to handle locals. Ruby is more dynamic than most other dynamic languages used in industry, so direct comparisons are challenging, but Lua's upvalue semantics are quite close to Ruby's escaping locals.

\subsection{JRuby}

JRuby is an implementation of Ruby for the JVM that compiles a custom IR to JVM bytecode, which may then be JIT-compiled by the JVM~\cite{Nutter18}.
Escape analysis is performed at IR construction and processes all locals in a scope as a single \emph{binding}.
All locals in a binding are pessimistically assumed to escape unless static analysis can prove otherwise.
If it is at all possible that any local in a binding \emph{could} be captured, or the binding itself could be reified (e.g., the presence of a block literal or a send to a dynamically named method), then that binding is marked as escaping.
Notably, that analysis is name-based, and does not consider aliases of or object references to methods known to escape, such as \texttt{eval}.

Once the JRuby IR is compiled, non-escaping bindings are never materialized; the locals are stored individually in the JVM stack frame's local variable array, which the JVM's JIT compiler can then elide, storing Ruby locals in machine registers.
Escaping bindings are placed in a heap-allocated environment\footnote{JRuby calls the environment a \emph{dynamic scope} with the $(\textsf{name}, \textsf{pos})$ stored in an associated \emph{static scope}.} with static links to parent environments, where they are accessed indirectly via a $(\textsf{depth}, \textsf{offset})$ pair.
There is no speculative optimization that stores potentially escaping locals in fast storage, thus there is no need to transition locals between storage locations.
However, that means potentially escaping locals that never escape at run time are always heap-allocated.

\subsection{TruffleRuby}

TruffleRuby~\cite{seaton2014, woess2014} is an implementation of Ruby for the GraalVM~\cite{duboscq2013graal}, built as a Truffle-based AST interpreter~\cite{wimmer2012, wuerthinger2012}, with a core library implemented in Ruby.
Unlike CRuby or JRuby, TruffleRuby does not perform its own escape analysis, instead relying on metacompilation for optimized code generation.

TruffleRuby uses the Truffle frame API, which represents a frame as a set of typed storage slots with an accompanying descriptor to define the frame layout.
As normal Java objects, frames can be passed between AST nodes for data access, and live on the heap during interpretation.
During partial evaluation, AST nodes are inlined into a single compilation unit, slot indices are constant-folded, and frame operations are transformed into specialized IR nodes to aid escape analysis.
Afterwards, partial escape analysis will scalar-replace non-escaping frames and convert accessed slots to SSA values.
Otherwise, escaped frames are materialized and chained by static links, much like JRuby's heap-allocated environments.
While reading locals from a materialized frame requires indirect access using a \texttt{(slot, depth)} locator, the slot type information is used to store locals more efficiently, avoiding boxing where possible.

Whereas JRuby treats all bindings that \emph{could} statically escape as escaped, TruffleRuby adapts to run-time behavior.
Deoptimization stubs are used for paths not yet taken, allowing partial escape analysis to ignore those paths.
Should the run-time behavior change such that a frame now escapes, it would trigger deoptimization and transfer control back to the interpreter, reconstructing the heap-allocated Truffle frame as it would any other value.
Subsequent recompilation will be with the refined view of the world and the frame will be materialized accordingly.

\subsection{LuaJIT}

Lua 5.1\footnote{Lua's variable resolution changed in 5.2, but we're limiting our consideration to the version supported by LuaJIT.} is a dynamic language that has locals, globals, and lexically resolved \emph{upvalues}~\cite{IerusalimschyFC05}.
Locals are introduced with the \texttt{local} keyword on variable assignment and will be stored on the VM's value stack.
Unlike Ruby, variables cannot be accessed through dynamically constructed names, and so the bytecode compiler can resolve all variable accesses statically.
If a name does not correspond to a local, search proceeds up through successive enclosing scopes until a match is found.
If no match is made, the variable is considered global and access goes through the globals table.
A non-local match is recorded in the current scope as an upvalue, which aliases the resolved variable's storage location.

Because upvalues can be captured in closures, they can escape their current scope.
A non-escaping\footnote{Lua uses the terms \emph{open} and \emph{closed}, rather than \emph{escaped}, but they're conceptually the same. We use \emph{escaped} to ease comparison with Ruby.} upvalue can alias the original stack slot.
When an upvalue escapes, the stack value is copied into a cell within the upvalue itself and the alias retargeted.
There is no mechanism for escaping an entire scope, so all escaped upvalue targets are moved separately.

LuaJIT is a tracing JIT compiler for Lua, compiling Lua bytecode to an SSA-based IR, and falls back to an optimized bytecode interpreter written in hand-crafted assembly~\cite{Pall09}.
A LuaJIT trace performs interprocedural analysis and transforms both local and non-escaping upvalue accesses within traced frames into SSA values, which can then be optimized away or turned into fast machine register access.
Escaped upvalues are accessed through their storage cell, although a read-only upvalue may still be constant-folded away.

Lua's metaprogramming facilities aren't as rich as Ruby's and that simplifies optimization.
E.g., Lua has a \texttt{load} operation, but not \texttt{eval}; \texttt{load} can compile a snippet of code but has no access to the enclosing scope.
Lua tables can be accessed with dynamic names, but scopes cannot.
There is no way to reify or retain a binding.
While the \texttt{debug} API can set values on the stack or an upvalue, there is no way to introduce a new local.

\subsection{LLVM}

LLVM, a modular compiler toolchain, implements local variables (``automatic
variables'') by first allocating them with \texttt{alloca} and then implementing
into-SSA in a pass called \texttt{mem2reg} that promotes memory locations to SSA
values~\cite{lattner2002llvmisa, lattner2002llvm, lattner2004llvm}.

This is simpler in languages such as C and C++, which do not offer support for
such flexible reflection as Ruby.

\section{Conclusion}

This paper introduced a new method-based JIT compiler for Ruby,
\zjit{}, and compared it with both the reference interpreter and the existing JIT
compiler, YJIT.

We discussed the semantics of local variables in Ruby and why they are
difficult to implement efficiently. We presented a new optimistic
assumption-based approach for compiling locals and the current bugs we are
fixing with it.

\zjit{} is currently faster than both the interpreter and YJIT on
microbenchmarks. It is also faster than the interpreter on, and competitive
with YJIT on, larger applications such as Rails-based web applications.

We do not currently have a partial-escape-analysis-based optimization pass for
further optimizing locals in blocks. We leave this for future work.

We also leave implementing many built-in library functions in Ruby as future
work. This will help the compiler analyze and optimize more code than it can
right now.

The source is available on GitHub in the \texttt{zjit/} directory of the upstream
CRuby repository, \url{https://github.com/ruby/ruby}.

\section*{Acknowledgements}

We thank the reviewers and Benoit Daloze for their excellent feedback. We thank
Maxime Chevalier-Boisvert for her work as a founding member of the \zjit{} team.
We thank last, but definitely not least, the open source contributors who have
spent time and energy sending us high-quality patches.

\bibliographystyle{ACM-Reference-Format}
\bibliography{refs}

\end{document}